\documentclass{PoS}

\title{Replica-Exchange Cluster Algorithm\thanks{
Work supported by the Deutsche Forschungsgemeinschaft (DFG)
under grant Nos.~JA483/22-1 and JA483/23-1, 
the EU RTN-Network `ENRAGE': {\em Random Geometry
and Random Matrices: From Quantum Gravity to Econophysics\/} under grant
No.~MRTN-CT-2004-005616, and by the computer-time grant No.~hlz10 of 
the John von Neumann Institute for Computing (NIC),
Forschungszentrum J\"ulich. EB thanks the DAAD for a travel grant.
}}

\ShortTitle{Replica-Exchange Cluster Algorithm}

\author{\speaker{Wolfhard Janke} and Elmar Bittner\\
        Institut f\"ur Theoretische Physik, Universit\"at Leipzig,\\ Postfach 100\,920, D-04009 Leipzig, Germany\\
        E-mail: \email{wolfhard.janke@itp.uni-leipzig.de}}

\abstract{
In finite-size scaling analyses of Monte Carlo 
simulations of second-order phase transitions one often needs
an extended temperature/energy range around the critical point. By
combining the replica-exchange algorithm with cluster updates and
an adaptive routine to find the 
range of interest, we introduce a new flexible and powerful method
for systematic investigations of critical phenomena.
As a result, we gain two further orders of magnitude in the performance for 2D and 3D Ising models
in comparison with the recently proposed Wang-Landau recursion for cluster algorithms
based on the multibondic algorithm,
which is already a great improvement over the standard multicanonical variant.
}

\FullConference{The XXVII International Symposium on Lattice Field Theory - LAT2009\\
		 July 26-31 2009\\
		 Peking University, Beijing, China}

\begin{document}


While much attention has been paid in the past to simulations of
first-order phase transitions and systems with rugged free-energy landscapes
in generalized ensembles (umbrella, multicanonical, Wang-Landau, 
parallel/simulated tempering sampling) \cite{rugged}, the merits of this 
non-Boltzmann sampling approach also for simulation studies of critical
phenomena has been pointed out only recently. In Ref.~\cite{bbwj}, Berg and 
one of the authors combined multibondic sampling \cite{JaKa95} with the
Wang-Landau recursion \cite{WaLa01} to cover the complete desired ``critical''
temperature range in a single simulation for each lattice size, where the 
``desired'' range derives from a careful finite-size scaling (FSS) analysis
of all relevant observables. Since the individual reweighting ranges of the
involved observables may be quite disparate, this scaling analysis is the second 
important ingredient of the method.

\begin{figure}[t]
\centering
\includegraphics[width=0.55\textwidth]{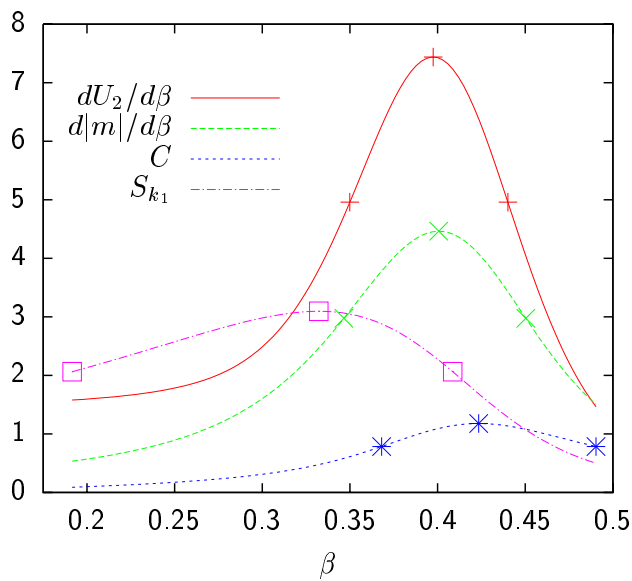}
\caption{\label{betafig}Reweighted observables for the 2D Ising model with $L=8$. The symbols mark the maximum values 
$S^{\rm max}_L$ and the value  $S_L(\beta^{+/-}_S)=r S^{\rm max}_L$ with $r=2/3$.}
\end{figure}

Our new replica-exchange cluster algorithm is a combination of 
parallel tempering methods~\cite{pt} with the Swendsen-Wang cluster algorithm~\cite{sw}. 
For the parallel tempering procedure we use a set 
of $N_{\rm rep}$ replica, where the number of replica depends on the reweighting range that is
needed for the FSS analysis~\cite{opti}. 
To determine this range we perform at the beginning of our simulations
a short run in a reasonable temperature interval.
We choose the number of replica $N_{\rm rep}$ so that the overlap of adjacent histograms is 
always larger than $25\%$. This is necessary to ensure that the multi-histogram reweighting~\cite{multi} 
works properly.
Using the data of this short run as input for the 
multi-histogram reweighting routine, we determine the pseudo-critical points $C_L^{\rm max}=C_L(\beta^{\rm max}_{C,L})$ of 
the specific heat $C(\beta)=\beta^2 V (\langle e^2\rangle - \langle e\rangle^2)$ and $\chi_L^{\rm max}$ of
the susceptibility $\chi(\beta)=\beta V (\langle m^2\rangle - \langle |m|\rangle^2)$, where
$e=E/V$ is the energy density and $m=M/V$ the magnetization density. Furthermore, we 
measured the maxima of the slopes of the magnetic cumulants, $U_2(\beta)=1-\langle m^2\rangle/3\langle |m|\rangle^2$ 
and $U_4(\beta)=1-\langle m^4\rangle/3\langle m^2\rangle^2$, and of the derivatives of the
magnetization, $d\langle |m|\rangle/d\beta$, $d\langle \ln |m|\rangle/d\beta$, and $d\langle \ln m^2\rangle/d\beta$, respectively. 
We also include the first structure factor $S_{k_1}$ (see, e.g., Ref.~\cite{stanly}) 
in our measurement scheme to be directly comparable with the results of Ref.~\cite{bbwj}.
Then we determine the $\beta$ values where the observables $S=\{C,\chi,\dots \}$ reach the
value $S_L(\beta^{+/-}_{S,L})=r S^{\rm max}_L$, where we use the moderate value $r=2/3$.  
This leads to a sequence of $\beta^{+/-}_{S,L}$ values, where $\beta^+_{S,L} > \beta^{\rm max}_{S,L}$ and $\beta^-_{S,L} < \beta^{\rm max}_{S,L}$. 
In Fig.~\ref{betafig}, we show as an example for such a sequence
the reweighted curves for $C$, $\chi$, $dU_2/d\beta$, 
and $S_{k_1}$ for the two-dimensional (2D) Ising model with linear lattice size $L=8$. 
The actual simulation range is then given by the largest interval of the sequence of all $\beta_{S,L}^{+/-}$ values. 
In our example in Fig.~\ref{betafig}, this would lead to an interval $[\beta_{S_{k_1},L}^-,\beta^+_{C,L}]$.
As one also can see in this figure, the value of $\beta_{S_{k_1},L}^-$ is further away from the critical point 
then all other $\beta_{S,L}^{+/-}$ values; therefore, if one is not particularly interested in the first structure factor,
then the simulation range can be chosen narrower. 
We now use the thus determined interval with the same number of replica for our actual measurement run. This interval is usually 
narrower then the original estimate, hence the overlap of the histograms becomes larger and the 
applicability of the multi-histogram reweighting method is assured. 

Let us now illustrate the work flow of our new method for the 2D Ising model.
Here we started with a reasonable choice of the temperature interval,
$\beta_8^-=0.15$ and $\beta_8^+=0.6$, for our smallest lattice $L=8$. For the large systems 
we used the measurement interval of the smaller system as input interval. Then we used the following
general recipe:

\begin{enumerate}
\item choose an input temperature interval
\item choose the number of replica
\item compute the simulation temperatures for the replica (e.g., equidistant in $\beta$ \cite{footnote1,katz})
\item perform several hundred thermalization sweeps
\item perform a short measurement run
\item check the overlap of the energy histograms: if the overlap is too small, add 2 replica and goto step 3, else go on 
\item use multi-histogram reweighting to determine $\beta_{S,L}^+$ and $\beta_{S,L}^-$ for all observables $S$
\item update the temperature interval according to the largest
interval of $\beta_{S,L}^+$ and $\beta_{S,L}^-$,\\ i.e., $[\min_S \{\beta_{S,L}^-\}, \max_S \{\beta_{S,L}^+\}]$ 
\item perform several hundred thermalization sweeps
\item perform the measurement run
\end{enumerate}

\begin{table}[t]
\caption{Simulation range and numbers of replica for the 2D Ising model simulations on $L^2$ lattices.\label{tab2d}}
\centering
\begin{tabular}{rrrr}
\hline \hline
\makebox[0.7cm][r]{$L$}& \makebox[1.8cm][c]{$\beta_L^{-}$} & \makebox[1.8cm][c]{$\beta_L^{+}$} & \makebox[0.7cm][r]{$N_{\rm rep}$} \\ \hline
8    & 0.194\,654 & 0.488\,895 & 4  \\
16   & 0.319\,082 & 0.469\,406 & 6  \\
32   & 0.380\,126 & 0.458\,969 & 6  \\
64   & 0.410\,836 & 0.452\,740 & 10 \\
128  & 0.425\,789 & 0.447\,917 & 10 \\
256  & 0.433\,297 & 0.444\,115 & 12 \\
512  & 0.436\,997 & 0.443\,161 & 12 \\
1024 & 0.438\,407 & 0.442\,653 & 16 \\ \hline 
$\infty$ &\multicolumn{2}{c}{$\beta_c=0.440\,686\,7935\dots$} &\\ \hline \hline

\end{tabular}
\end{table}

\begin{figure}[t]
\centering
\includegraphics[width=0.55\textwidth]{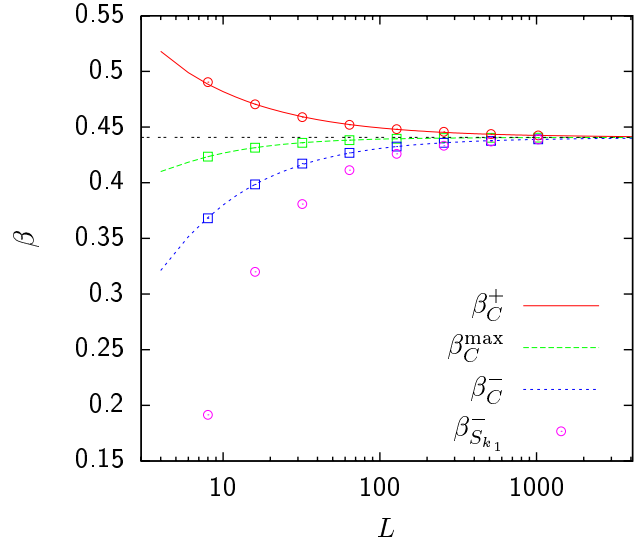}
\caption{The temperature interval determined using the specific heat as a function of the system size.
The horizontal line indicates the critical inverse temperature, the other lines show the exact results calculated from the
formula of Ferdinand and Fisher~[12]. The	circles indicate the
simulation ranges determined fully automatically, cf. Table~1, and the boxes show
for completeness the measured values for $\beta_C^{\rm max}$ and $\beta_C^-$.  \label{fig_c_beta} }
\end{figure}

After choosing an input temperature interval and a number of replica for the smallest system, our program
simulated system sizes from  $L=8$ up to $L=1024$ fully automatically. This shows how robust our new method is.
Table~\ref{tab2d} gives an overview of the automatically determined
temperature intervals, which roughly scale with $L^{-1/\nu}$, where $\nu$ is the standard critical exponent of the correlation length.
This scaling can also be used to extrapolate the input interval for larger system sizes.
In two dimensions, the branch coming from the specific heat 
has a logarithmically scaling, therefore, one could use this knowledge
to improve the extrapolation for this special case. We refrain from
such modifications to keep the program as generally usable as
possible.
For comparison we show in Fig.~\ref{fig_c_beta} the calculated
temperature interval $[\beta_{C,L}^-,\beta^+_{C,L}]$ using the
specific heat formula of Ferdinand and Fisher~\cite{FF} and the automatically determined
interval of our algorithm.

\begin{figure}[t]
\centering
\includegraphics[width=0.55\textwidth]{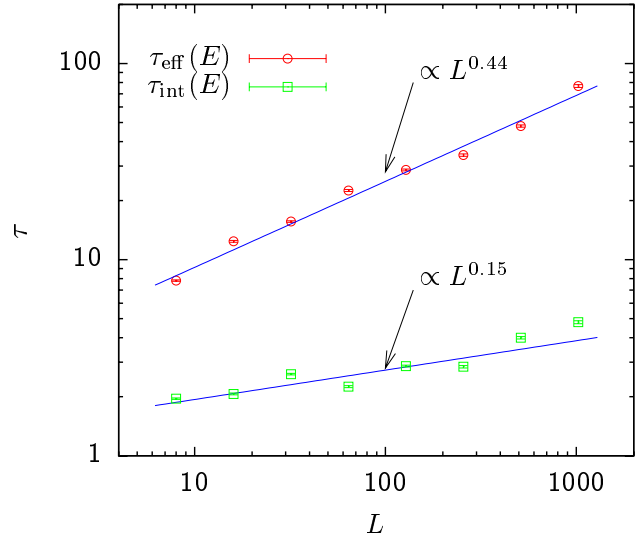}
\caption{\label{tau2d}Autocorrelation times $\tau_{\rm int}$ and $\tau_{\rm eff}$ for the energy of the 2D Ising model, where $\tau_{\rm eff}=N_{\rm
rep}\times
\tau_{\rm int}$ and $N_{\rm rep}$ is the
number of replica.}
\end{figure}

To assess the performance of the method, we measured the integrated autocorrelation time $\tau_{\rm int}$ for each temperature 
and system size. We found that
the integrated autocorrelation times $\tau_{\rm int}$ for the energy, squared magnetization and the first structure factor
scale only weakly with the system size $L$. As an example we show $\tau_{\rm int}(E)$ as a function of $L$ in Fig.~\ref{tau2d}.
The critical slowing down scales $\propto L^z$, here we find a dynamical critical exponent $z=0.15(3)$. 
When we also take the number of replica $N_{\rm rep}$ into account and define an effective autocorrelation times $\tau_{\rm eff}$ according 
to $\tau_{\rm eff}=N_{\rm rep}\times \tau_{\rm int}\propto L^{z_{\rm eff}}$, we find a power law with an exponent
$z_{\rm eff}=0.44(2)$.
For $\tau_{\rm int}$ and $\tau_{\rm eff}$ of $m^2$ we find slightly smaller values, $z=0.09(3)$ and $z_{\rm eff}=0.37(3)$, respectively. 
For $S_{k_1}$ the dynamical critical exponent is compatible with $0$ and for $\tau_{\rm eff}$ we find $z_{\rm eff}=0.29(2)$.
Even the larger absolute values of the effective autocorrelation times are almost an order of magnitude smaller
and scale with a much smaller exponent than  using the recently proposed multibondic Wang-Landau method~\cite{bbwj}.

\begin{figure}[t]
\centering
\includegraphics[width=0.55\textwidth]{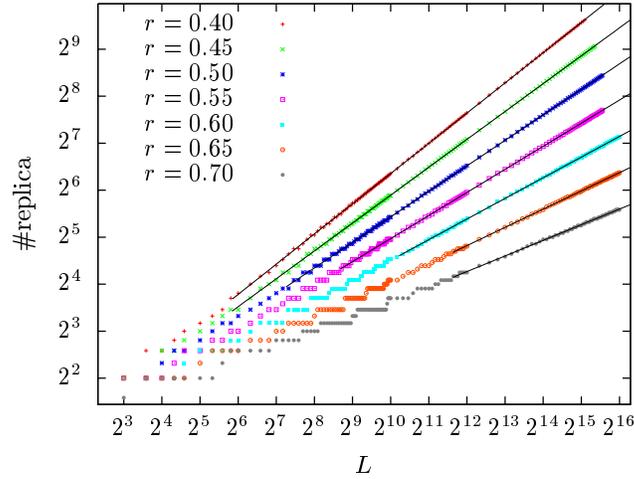}
\caption{\label{N_scal} The number of replica needed to cover the
range of interest for the specific heat plotted as a
function of the system size for the 2D Ising model. 
The straight lines are least square fits according to $const.\,L^{(1-r)/\nu}$, 
with $r$ and $\nu=1$ fixed. 
}
\end{figure}

Due to the fact that in the 2D Ising model the critical exponent $\alpha$ of the specific heat is zero,
the reweighting range of a single Monte Carlo simulation is $\propto L^{-1/\nu}$ whereas
the range of interest scales with $L^{-r/\nu}$ (with the range parameter $r$ defined above and $\nu=1$). 
The number of replica needed thus increases with the system size as
$L^{(1-r)/\nu}$, cf. Ref.~\cite{bbwj1}. In Fig.~\ref{N_scal} we 
show the numbers of replica needed as a function of the system size for various values of $r$. We also included least square fits 
according to the previously given scaling form and find a reasonable agreement. 
In the 3D Ising model where $\alpha  \approx 0.11$, the 
reweighting range scales equally to the range of interest according to $L^{-1/\nu}$, so that we
can use here the same number of replica for all system sizes.
In our 2D simulations only the $\beta^+_L$ branch is determined by the
scaling of the specific heat. If one omits $C$ as a criterion to specify the range
of interest in this non-generic case, the numbers of replica can also be fixed for all system
sizes. As a nice side effect, the dynamical critical exponent for the effective autocorrelation  times $\tau_{\rm eff}$ is
even smaller than in the case including $C$, we find $z=z_{\rm eff}=0.32(1)$ for the energy and $z=z_{\rm eff}=0.24(1)$ for $m^2$. If the
reweighting range is now too narrow to determine the critical exponent $\alpha$ directly, one still can use the hyperscaling relation $\alpha=2-D\nu$ with 
the dimensionality $D$.

\begin{figure}[t]
\centering
\includegraphics[width=0.55\textwidth]{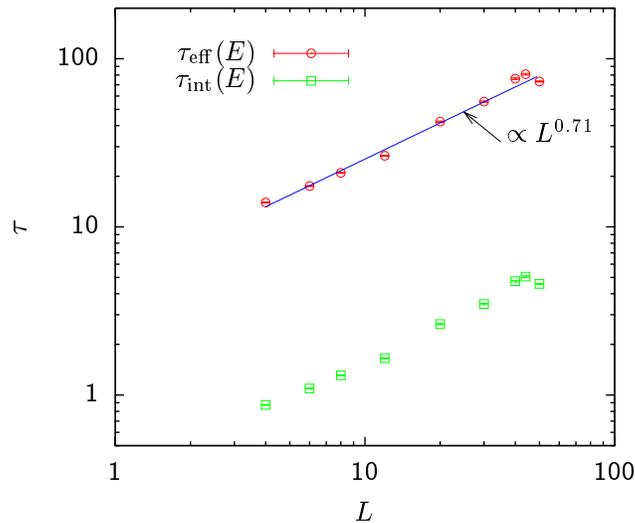}
\caption{\label{tau3d}$\tau_{\rm int}$ and $\tau_{\rm eff}$ for the 3D Ising model, where 
$\tau_{\rm eff}=N_{\rm rep}\times \tau_{\rm int}$ and $N_{\rm rep}=16$ is the number of replica.}
\end{figure}

In Table~\ref{tab3d} we give an overview of the automatically determined temperature intervals 
for the 3D Ising model which are similar to the intervals compiled in Table~I of Ref.~\cite{bbwj}.
By increasing the numbers of sweeps in the first short measurement run would lead to a better estimate for the
temperature interval. We used only about $1\%$ of our CPU time for this determination, increasing this
percentage may gain a further improvement of the final results.
In Fig.~\ref{tau3d} we show the integrated autocorrelation times as well as the
effective autocorrelation times for the energy of the  3D Ising model. Here we find for 
the dynamical critical exponent $z=0.71(3)$. 
In this case $\tau_{\rm eff}$ is just a constant shift for all system sizes, 
due to the fact that the number of replica is independent of the system size.
We find $z=0.70(3)$ for the autocorrelation times of $m^2$ and $z=0.38(4)$ for $S_{k_1}$.   
In the 3D Ising model the absolute values of the integrated autocorrelation times are 
almost two orders of magnitude smaller and even the effective autocorrelation times
are an order of magnitude smaller than those reported for the multibondic scheme in Ref.~\cite{bbwj}. 
Since also the dynamical critical exponents are smaller, the asymptotic critical slowing down is less pronounced.

To summarize, we have introduced a very flexible approach for a systematic determination and simulation of
the critical energy range of interest for second-order phase transitions, 
which one needs to measure critical exponents. 
The efficiency of the method depends of course on the chosen or available update scheme (Metropolis, heat-bath,
Glauber, cluster, $\dots$) in the particular case.  
Since our method is completely general and can be used with any update scheme, it could be employed for
all simulations in high-energy physics and quantum field theory, statistical physics, chemistry and biology where one is
interested in critical exponents.

\begin{table}[t]
\caption{Simulation range and numbers of replica for the 3D Ising model simulations on $L^3$ lattices.\label{tab3d}}
\centering
\begin{tabular}{rrrr}
\hline \hline
\makebox[0.7cm][r]{$L$}& \makebox[1.8cm][c]{$\beta_L^{-}$} & \makebox[1.8cm][c]{$\beta_L^{+}$} & \makebox[0.7cm][r]{$N_{\rm rep}$} \\ \hline
20 & 0.211\,098 &  0.233\,487 & 16 \\
30 & 0.216\,204 &  0.228\,823 & 16 \\
44 & 0.218\,717 &  0.226\,695 & 16 \\
56 & 0.219\,651 &  0.225\,533 & 16 \\
66 & 0.220\,115 &  0.224\,196 & 16 \\
80 & 0.220\,517 &  0.224\,195 & 16 \\ \hline \hline
\end{tabular}
\end{table}

\begin{thebibliography}{99}
\bibitem{footnote1}
The method can presumably be further
improved by optimizing the simulated temperature set by using the feedback-optimized parallel tempering MC~\cite{katz}.
Due to the fact that the replica are already very close to each other, the improvement, however, might be smaller than
the additional computational effort to determine the feedback-optimized temperature set.
\end{thebibliography}
\end{document}